\documentstyle[12pt]{article}
\begin{document}


\title{The chiral magnetic effect in hydrodynamical approach}

\author{A.V. Sadofyev\footnote{sadofyev@itep.ru}, M.V. Isachenkov\footnote{m.isachenkov@gmail.com}\\
\emph{Institute of Theoretical and Experimental Physics, Moscow}}
\date{}

\maketitle

\begin{center}{\bf Abstract}\end{center}
We consider a model with two conserved currents, vector and axial, and two
associated chemical potentials, $\mu$ and $\mu_5$. In the presence
of external magnetic and electric fields the axial current is anomalous.
Generalizing recent results and using thermodynamic relations alone we
demonstrate that one can evaluate the chiral magnetic effect (ChME), that is
the electric current flowing in the direction of the external magnetic
field. In the linear in the chemical potential approximation
the current is the same as for non-interacting fermions. In other words,
there exists a hydrodynamic
“non-renormalization theorem for the ChME”.

\section{Introduction}
The chiral magnetic effect
(for review and references see \cite{kharzeev1})
gradually becomes a litmus-paper for quark-gluon plasma physics.
So much the more interesting is the task of its explanation and study.
The effect itself is the generation of
electric current  by applying an external magnetic field
{\bf H} to a chiral medium:
 \begin{equation}
 {\bf j}_{el}~=~c_{ch}{\bf H}~,
 \end{equation}
 where $c_{ch}$ is a constant.
 The problem is to evaluate the coefficient $c_{ch}$.
 In particular, for fermions interacting with the magnetic field
 but not among themselves one gets \cite{kharzeev1},\cite{kharzeev2}:
 \begin{equation}\label{one}
 c_{ch}~=~N_c \frac{e^2 \mu_5}{2\pi^2}~,
 \end{equation}
 where $N_c$ is the number of colors, $e^2$ is the electric charge squared and
 $\mu_5$ is the chemical potential conjugated to the the difference in the
 number of left- and right-handed fermions. Taking into account the
 interaction between the fermions is a challenge.

  In our considerations, we start from the hydrodynamic  approximation which
  is, in a way, opposite to the limit of  free fermions. Indeed, in
  the hydrodynamic approximation one considers distances $l$ much larger than
  the free path of the particles of the medium,
  $$l~\gg~ l_{free~path}~~.$$
  The hydrodynamic approach to chiral media is pioneered in \cite{son} where it is demonstrated that
  the general thermodynamic identities are powerful enough to fix some transport
  coefficients.  In the model of \cite{son}
   all the currents are chiral.
   We generalize these considerations to the case of vector and axial currents separately conserved (at least in the
   absence
   of external electromagnetic fields). The  two-current set-up (see also \cite{Gao}) enables physical quantities to have their proper parity.

With all the calculations completed, we’ll obtain the first term in
the  expansion of the coefficient $c_{ch}$ in the axial chemical potential $\mu_5$, whereas in coupling constant it  includes all orders due to our use of  the hydrodynamical approximation. 
The result coincides with (\ref{one}).
The fact that calculated coefficient coincides with those derived in linear approximation permits us to formulate a “non-renormalization theorem” for ChME that seems to be analogous to already existing non-renormalization theorems \cite{Adler}: the ChME coefficient doesn’t receive corrections in coupling constant, but only in chemical potentials of right- and left-handed particles. In fact, this theorem is a consequence of the second law of thermodynamics.

Through the paper we use the following conventions: the metric tensor has (-,+,+,+) signature, 4-velocity of an element of the liquid satisfies $u_\mu u^\mu=-1$, the quantities which correspond to the electric and magnetic fields in the rest frame are $E^\mu=F^{\mu\nu}u_\nu$ and $B_\mu=\frac{1}{2}\epsilon_{\mu\nu\alpha\beta}u^\nu F^{\alpha\beta}$. We also introduce the
following notation:$$\omega_\mu=\frac{1}{2}\epsilon_{\mu\nu\alpha\beta}u^\nu \partial^\alpha u^\beta~.$$ For   simplicity we consider an Abelian case henceforth ($N_c=1$).

\section{Hydrodynamical approximation}

We discuss a hydrodynamical model of the ideal chiral liquid   in the presence of the chiral anomaly. To describe a nonzero chirality we introduce independent densities of the right- and left-handed particles or, equivalently,   density $n$ and "axial" density $n_5$ (which phenomenologically correspond to the QCD currents $J^{\mu}=\bar q \gamma^{\mu} q$ and $J^{\mu}_5=\bar q \gamma^{\mu} \gamma^5 q$). We also introduce their chemical potentials $\mu,\mu_5$. This model is described by following equations:

\begin{eqnarray}
\partial_{\mu}T^{\mu\nu}=F^{\nu\lambda}j_{\lambda}\nonumber\\
\partial_{\mu}j^{\mu}=0~~~~~~~~~\nonumber\\
\partial_{\mu}j_5^{\mu}=CE^{\mu}B_{\mu},
\end{eqnarray}

\noindent where the stress-energy tensor, vector and axial current are

\begin{eqnarray}
T^{\mu\nu}=w u^{\mu}u^{\nu}+p g^{\mu\nu}\nonumber\\
j^{\mu}=nu^{\mu}+\nu^{\mu}~~~~~~~~~\nonumber\\
j_5^{\mu}=n_5 u^{\mu}+\nu_5^{\mu}~~~~~~~.
\end{eqnarray}

\noindent Terms $\nu,\nu_5$ are higher-order corrections in velocities and their derivatives, which corresponds, in particular, to dissipative effects. There is no a dissipative force in the rest frame of the element of a liquid. Therefore $\nu,\nu_5$ satisfy conditions(see \cite{landau}):

\begin{eqnarray}
\nu_{\mu}u^{\mu}=0\nonumber\\
\nu_{5,\mu}u^{\mu}=0.
\end{eqnarray}

Following the pattern of \cite{son}, let us consider
 consequences from the second law of thermodynamics. Transforming $u_\nu \partial_\mu T^{\mu\nu}+\partial_\mu j^\mu + \partial_\mu j_5^\mu$ we get:

\begin{eqnarray}
\partial_\mu(su^\mu-\frac{\mu}{T}\nu^\mu-\frac{\mu_5}{T}\nu_5^\mu)=-\nu^\mu\partial_{\mu}\frac{\mu}{T}-\nu_5^\mu\partial_{\mu}\frac{\mu_5}{T}+\frac{E^\lambda}{T} \nu_{\lambda}-\mu_5 \frac{C}{T} E\,B.
\end{eqnarray}

\noindent The second law of thermodynamics implies that the entropy is non-decreasing and the entropy current divergence is non-negative. The entropy current is usually introduced as:

\begin{eqnarray}
s^\mu=su^\mu-\frac{\mu}{T}\nu^\mu-\frac{\mu_5}{T}\nu_5^\mu.
\end{eqnarray}

\noindent For the ideal liquid $\nu$ and $\nu_5$ are zero. But in the presence of the anomaly such a choice does not guarantee fulfilment of the second law of thermodynamics since the
 $CE\cdot B$ term can have both signs.  To compensate for the contribution of this term we can introduce in $\nu$,in $\nu_5$ and in $s_\mu$ further terms,  not forbidden by the second law of thermodynamics in the presence of the anomaly. We choose such terms in the accordance with \cite{son}:

\begin{eqnarray}
s^\mu=su^\mu-\frac{\mu}{T}\nu^\mu-\frac{\mu_5}{T}\nu_5^\mu+D\omega^\mu+D_B B^\mu,
\end{eqnarray}

\begin{eqnarray}
\nu^\mu=\kappa\omega^\mu+\kappa_B B^\mu\nonumber\\
\nu_5^\mu=\xi\omega^\mu+\xi_B B^\mu.
\end{eqnarray}

\noindent Coefficients $\xi,\xi_B,\kappa,\kappa_B,D,D_B$ are functions of the thermodynamical variables. The coefficient in front of $B^\mu$ in the vector current corresponds to the ChME in hydrodynamical approach. To impose the second law of thermodynamics we demand the contributions to the divergence of the new terms to be compensated by the chiral anomaly. When $B^\mu,\omega^\mu,E\,\omega,E\,B$ are arbitrary, the following set of equations takes place:

\begin{eqnarray}
\label{s1}
(\partial_\mu D-2\frac{\partial_{\mu}P}{w}D-\kappa\partial_\mu\frac{\mu}{T}-\xi\partial_\mu \frac{\mu_5}{T})\cdot\omega^\mu=0\\
\label{s2}
(\partial_\mu D_B-\frac{\partial_{\mu}P}{w}D_B-\kappa_B\partial_\mu\frac{\mu}{T}-\xi_B\partial_\mu\frac{\mu_5}{T})\cdot B^\mu=0\\
\label{s3}
(\frac{2nD}{w}-2D_B+\frac{\kappa}{T})\cdot(E_\mu\omega^\mu)=0\\
\label{s4}
(\frac{n D_B}{w}+\frac{\kappa_B}{T}-\mu_5\frac{C}{T})\cdot(E_\mu B^\mu)=0.
\end{eqnarray}

\noindent To solve this set of equations one should use initial conditions. For this purpose it is convenient to change a thermodynamical variables from ($\mu$,$\mu_5$,$T$) to ($\bar\mu=\frac{\mu}{T}$, $\bar\mu_5=\frac{\mu_5}{T}$, $P$). Then

\begin{eqnarray}
dT=\frac{T}{w}dP-\frac{nT^2}{w}d\bar\mu-\frac{n_5 T^2}{w}d\bar\mu_5.
\end{eqnarray}

\noindent We can set the gradients of thermodynamical variables ($\partial P, \partial \frac{\mu}{T}, \partial \frac{\mu_5}{T}$) arbitrary at initial time. Then coefficients 
 in front of them must be zero at arbitrary time to make (\ref{s1}-\ref{s4})
 valid :

\begin{eqnarray}
\frac{\partial D}{\partial p}-\frac{2D}{w}=0~~,~~\frac{\partial D_B}{\partial p}-\frac{D_B}{w}=0\\
\frac{\partial D}{\partial \bar\mu}=\kappa~~~~~,~~~~~\frac{\partial D_B}{\partial \bar\mu}=\kappa_B~~~~~\\
\frac{\partial D}{\partial \bar\mu_5}=\xi~~~~~,~~~~~\frac{\partial D_B}{\partial \bar\mu_5}=\xi_B.~~~~
\end{eqnarray}

\noindent We search for a solution in the form:

\begin{eqnarray}
D=T^2 d(\bar\mu,\bar\mu_5)~~,~~D_B=T d(\bar\mu,\bar\mu_5)\nonumber~~~~\\
\kappa=\frac{\partial}{\partial\bar\mu}(T^2 d)~~~~,~~~~~\xi=\frac{\partial}{\partial\bar\mu_5}(T^2 d)~~~~\nonumber\\
\kappa_B=\frac{\partial}{\partial\bar\mu}(T d_B)~~~~,~~~\xi_B=\frac{\partial}{\partial\bar\mu_5}(T d_B),
\end{eqnarray}

\noindent where $d$ and $d_B$ are arbitrary functions. One finds after straightforward calculations:

\begin{eqnarray}
d_B=C\bar\mu_5\bar\mu+f(\bar\mu_5)~~,~~d=C\bar\mu_5 \bar\mu^2+\bar\mu f(\bar\mu_5)+F(\bar\mu_5),
\end{eqnarray}

\noindent where $f$ and $F$ are arbitrary functions.

 We are interested only in the ChME term which corresponds to the coefficient in front of $B_\mu$. For this coefficient $k_B$ in the lowest order in chemical potentials one gets:

\begin{eqnarray}
\kappa_B=C\mu_5+O(\mu_5^2,\mu\mu_5,\mu^2).
\end{eqnarray}

\noindent Higher-order terms in  $k_B$ are not fixed by this calculation. It's worth pointing out that the coefficients in front of $\omega^\mu$ have uncertainty even in the lowest order while in \cite{son} coefficient before $\omega^\mu$ in the chiral current is
uniquely fixed.

It is readily seen that there is a term proportional to the magnetic field in the vector current. This term corresponds to the ChME in the hydrodynamical model. Substituting the anomalous coefficient in front of $(E\cdot B)$ by $\frac{e^2}{2\pi^2}$ we finally get in the lowest order:

\begin{eqnarray}
\triangle j^\mu=\frac{e^2\mu_5}{2\pi^2} B^\mu.
\end{eqnarray}

\noindent The answer for an arbitrary $N_c$ is obtained by simple multiplication.

\section{Conclusion}
To summarize, it follows from the consideration of the entropy that
 the chiral magnetic effect is not vanishing  in
the hydrodynamical approximation. The corresponding coefficient $c_{ch}$, see
(\ref{one}), is expandable in the chemical potentials.
The first, linear term is uniquely fixed and 
 coincides with the result for non-interacting particles.
  Unlike the free-particle case, further terms
   in the expansion are not vanishing, generally speaking.
   They seem to be not fixed, however, by the general thermodynamic relations.
 The hydrodynamical approximation addresses physics in the infrared region of energies 
 and is sensitive, generally speaking, to all orders of perturbation theory.   Our result  
 suggests, therefore, that there exists  a new non-renormalization theorem for the coefficient
 $c_{ch}$ in the linear in the chemical potential approximation.  %
Besides the result itself which looks quite nonpresumable, one could wonder about power of pure thermodynamics in deriving of highly non-trivial relations in the field theory.

\section{Acknowledgments}
We are grateful to E. T. Akhmedov,  F. Gubarev and V. I. Zakharov  for valuable discussions.
The work of A. V. Sadofyev was supported by the grant for the Leading Scientific School NSh-6260.210.2 and DAAD Leonhard-Euler-Stipendium 2010-2011.


\begin{thebibliography}{99}

\bibitem{kharzeev1}
K. Fukushima,  , D. E. Kharzeev, H. J. Warringa, Phys. Rev. D {\bf 78}, 074033 (2008).

\bibitem{kharzeev2} D. E. Kharzeev, H. J. Warringa, Phys. Rev. D {\bf 80}, 034028 (2009).

\bibitem{son} D. T. Son,  P. Surowka, Phys. Rev. Lett. {\bf 103}, 191601 (2009).

\bibitem{Gao} S. Pu, J. Gao, Q. Wang, arXiv:1008.2418v2 [nucl-th].

\bibitem{Adler}
S. L. Adler, W. A. Bardeen, Phys. Rev. {\bf 182} (1969). 

\bibitem{landau} L. D. Landau and E. M. Lifshitz, \emph{Fluid Mechanics}, Pergamon, New York (1959).

\end{thebibliography}
\end{document}